\documentclass{pasa}%

\title[Scale-length of the Thin Disc]{Solar Space Density of the Red Clump Stars and the Scale-length of the Thin Disc}
\author[Yaz Gokce et al.]{E. Yaz G\"ok\c ce$^1$\thanks{esmayaz@istanbul.edu.tr}, S. Karaali$^1$, \c S. Duran$^2$, S. Bilir$^1$, A. Yal\c c\i nkaya$^3$, S. Ak$^1$, T. Ak$^1$, M. L\'{o}pez-Corredoira$^{4,5}$, A. Cabrera-Lavers$^{4,5}$\\
	\affil{$^1$Istanbul University, Faculty of Science, Department of Astronomy and Space Sciences, 34119, Beyaz\i t, Istanbul, Turkey} %
	\affil{$^2$Istanbul University, Graduate School of Science and Engineering, Department of Astronomy and Space Sciences, 34116, Beyaz\i t, Istanbul, Turkey} %
    \affil{$^3$Mithatpa\c sa cd. 16/4, 35320, Narl\i dere, \.Izmir, Turkey} %
    \affil{$^4$Instituto de Astrof\'{i}sica de Canarias, E-38205 La Laguna, Tenerife, Spain} %
    \affil{$^5$Departamento Astrof\'{i}sica, Universidad de La Laguna, E-38206 La Laguna, Tenerife, Spain}}%
\jid{PASA}
\doi{10.1017/pas.\the\year.xxx}
\jyear{\the\year}
\usepackage{natbib}
\begin{document}%

\begin{abstract}
We estimated the scale-length of the thin disc with the {\em J} and {\em W1} magnitudes of the most probable Red Clump (RC) stars in the Galactic plane, $-0\overset{^\circ}.5 \leq b \leq +0\overset{^\circ}.5$, in 19 equal sized fields with consecutive Galactic longitudes which cover the interval $90^\circ \leq l \leq 270^\circ$. Our results are constrained with respect to the solar space density ($D^*=5.95$), which indicates that the radial variation of the density is lower for higher Galactocentric distances. The scale-length of the thin disc is 2 kpc for the fields in the Galactic anticentre direction or close to this direction, while it decreases continuously in the second and third quadrants reaching to a lower limit of $h$ = 1.6 kpc at the Galactic longitudes $l$ = 90$^\circ$ and $l$ = 270$^\circ$. The distribution of the scale-length in 19 fields is consistent with the predictions from the Galaxia model and its variation with longitude is probably due to the inhomogeneity structure of the disc caused by the accreted material or other features such as warp and flare.
\end{abstract}

\maketitle%

\section{Introduction}
The photometrical and spectroscopical data obtained from the systematical sky survey programs have an important impact on the investigation of the structure, formation and evolution of the Galaxy. By the beginning of 2000’s, the data from the near-infrared sky surveys have contributed to the studies of the Galactic disc; for instance, the relation between the star formation regions and spiral arms, distribution of open clusters through the Galactic disc, formation and evolution of the open clusters, three-dimensional distribution of the interstellar matter in the Galactic disc, spatial distribution of the stars from different spectral types. 

Estimation of the Galactic model parameters, such as the scale-length, the scale-height, and the relative space densities of the Galactic populations, is a part of the investigation of the Galactic structure. The researchers used different methods for estimation of the Galactic model parameters for the thin disc, thick disc, and halo of our Galaxy. There has been an improvement for numerical values of the model parameters over 30 years, starting from the work of  \cite{G&R83}. We present those for only the thin disc in Table 1. There are many striking points for the numerical values of the Galactic model parameters for a given Galactic component, thin and thick discs and halo, as stated in the following: a) the range of values is large. For example, \cite{Chen01} and \cite{Siegel02} give the local space density of the thick disc as $6.5 - 13.5$ \% and $6 - 10$ \%, respectively. (b) the Galactic model parameters are absolute magnitude dependent \citep{Karaali04,Bilir06a}, (c) the Galactic model parameters are Galactic longitude/latitude dependent \citep{Buser98,Buser99,Bilir06b,Bilir06c,Bilir08,Cabrera07a,Ak07,YK10}, (d) the Galactic model parameters change with limiting distance of completeness \citep{Karaali07}. We should add that the scale-length also depends on the wavelength (cf. \citealp{Beckman96,Tuffs04}).

The observed features cited above can be explained by disc flares and warps. The disc of our Galaxy is not radially smooth and uniform. While the warp bends the Galactic plane upwards within the longitudes $0^\circ < l \leq 180^\circ $ and downwards within the longitudes $ 180^\circ < l \leq 360^\circ$, the flare changes the scale-height as a function of radial distance. Warp has been detected in the dust, gas and star components of the Galaxy \citep{Drimmel2001,Lopez02,Momany06,Marshall2006}. The stellar and gaseous flares are also consistent with each other \citep{Momany06,Kalberla14}. 

The overdensity regions also do affect the numerical values of the Galactic model parameters for a given Galactic component, thin and thick discs, and halo. This is the case for very large fields such as in the work of \cite{Juric08}, $6500$ deg$^2$. Such fields are contaminated by remnants of mergers which cause different numerical values for a specific model parameter, relative to the ones estimated in a relatively homogeneous region of the Galaxy. However, one should pay attention in explaining the origin of the overdensity regions in the Galaxy, i.e. such a region may originate either from a merger or from the affect of a warp of the Galaxy. The Canis Major overdensity region can be given as a typical example. \cite{Martin04} interpreted the overdensity region with Galactic coordinate $(l, b) = (240^\circ, -7^\circ)$ as the core of a satellite galaxy. However, \cite{Momany06}, \cite{Lopez06} and \cite{Lopez07} confirmed its coincidence with the Southern stellar maximum warp occurring near $l \sim240^\circ$ and $R \sim7$ kpc.

The red clump (RC) stars can be used for the investigation of the galactic structure. RC stars are core helium-burning giants. Their position in the colour-magnitude diagram (CMD) of open clusters as well as globular clusters is discrete relative to the main-sequence stars and red giants of these clusters. RC stars are also abundant in the solar neighbourhood which provide accurate absolute magnitudes due to their parallaxes. The small scattering of this population in the CMD has been used as an advantage to adopt a mean absolute magnitude for them in different optical and infrared (IR) bands. The final aim of the researchers was to use the mean absolute magnitude in distance estimation; could the RC stars be used as distance indicators? 

\cite{K&B99} claimed that the absolute magnitudes in the optical range of the RC stars within the spectral range of G8 III - K2 III lie within $0.7 \leq M_V \leq 1$ mag. Also, we should note that the \cite{Al00} sample covered a metallicity range of $-0.8 < [Fe/H] < +0.2$ dex. The absolute magnitude computed by \cite{G&S02} in the same band based on 14 open clusters with metallicity and age restrictions is compatible with the one of \cite{Al00}. 

An absolute magnitude in any optical or IR band free of metallicity and age would be the ideal one for the researchers. This was the case for the $I$ band in \citet{Udalski98}. However, a modest variation in $M_I$ with colour and metallicity has been claimed in the literature \citep{P&S98,S&G98,Sarajedini99,Zhao01,Kubiak02}. The dependence of the $M_I$ absolute magnitude on metallicity has been confirmed also theoretically \citep{S&G02}.

The numerical values claimed that the absolute magnitude $M_{K_{s}}$ in recent works are compatible with the former ones. \cite{H&G07} showed that $M_{K_{s}}=-1.57\pm0.05$ mag can be used for the determination of the distances of the open clusters with metallicities $- 0.5 < [Fe/H] \leq + 0.4$ dex and age $0.31 \leq t \leq 7.94$ Gyr. A value slightly different than the one just cited is the one in \cite{Groen08}, i.e. $M_{K_{s}}=-1.54\pm0.03$ mag, which is based on the re-reduced {\it Hipparcos} catalogue \citep{vanLee07}. \cite{Groen08} determined also the IR absolute magnitude for this sample as $M_I=-0.22\pm0.03$ mag. 

There are even more recent works related to the absolute magnitude of RC stars. \cite{Laney12} determined the absolute magnitude $M_{K_{s}}=-1.613 \pm 0.015$ mag for RC stars in the solar neighbourhood and applied it to estimate the distance of the Large Magellanic Cloud. \cite{KBY13} fitted the $M_{K_{s}}$ absolute magnitudes of the RC stars to a linear polynomial in terms of $V - K_s$ colour, and \cite{Bilir13a} calibrated the $M_{K_{s}}, M_V, M_J$ and $M_g$ absolute magnitudes in terms of colours corresponding to the photometric systems where the $K_s, V, J$ and $g$ bands are defined. Finally, we quote the paper of \cite{Bilir13b} where the $M_V$ absolute magnitude had been calibrated in terms of metallicity and $B - V$ colour. The data are based on a large sample of open and globular clusters with a wide range of metallicity. A mean value for the $M_{K_{s}}$ absolute magnitude with weak metallicity dependence can be used for distance estimation which is rather useful for the Galactic structure and interstellar extinction. Examples can be found in \cite{Lopez02,Lopez04}, \cite{Cabrera05,Cabrera07a,Cabrera07b,Cabrera08} and \cite{Bilir12} and references therein. 

The absolute magnitudes of the RC stars has been investigated also with the Wide-field Infrared Survey Explorer ({\it WISE}; \citealp{Wright10}). \cite{Yaz13} used the RC stars in the solar neighbourhood and determined a mean absolute magnitude on each band of $W1$ (3.4 $\mu$m) and $W3$ (12 $\mu$m) as well as a mean colour, i.e. $M_{W_{1}}=-1.635\pm0.026$, $M_{W_{3}}=-1.606\pm0.024$, and $(W1-W3)_0=-0.028\pm0.001$ mag. {\it WISE} can go at least one magnitude deeper than the Two Micron All Sky Survey (2MASS; \citealp{Skrutskie06}). Also, {\it WISE} can be affected from the Galactic extinction less than the other photometric systems, including 2MASS. Hence, one expects more accurate results for a larger sample of RC stars investigated with {\it WISE}.

In this paper we estimate the scale-length of the thin disc in the Galactic plane. We used the RC stars with Galactic latitude $-0\overset{^\circ}.5 \leq b \leq +0\overset{^\circ}.5$ and Galactic longitude $90^\circ \leq l \leq 270^\circ$ and evaluated the space density functions for equally spaced 19 fields. The Galactocentric radial distance range in our fields is 8 $\le R \le$ 16 kpc. Additionally, the Galactic latitude range, $-0\overset{^\circ}.5 \leq b \leq 0\overset{^\circ}.5$, is rather small. Hence, we considered the flare as negligible and fitted the observational space density profiles to a Galaxy model which involves only the exponential factor with scale-length. The evaluation of the solar space density of the RC stars by using the {\it Hipparcos} catalogue \citep{vanLee07} is given in Section 2. Section 3 is devoted to the estimation of scale-length, and finally a discussion is presented in Section 4.

\begin{table*}
	\caption{Scale-height and scale-length values of thin disc in the literature.} 
	\begin{center}
		\begin{tabular}{ccccl}
			\hline \hline
		Data	&	Object	&	$H$	&	$h$	  &	Reference\\
		        &		        &(pc)	& (kpc) & \\
		        \hline	
		  $BVI$	&	dwarf	        &	325	&	4.0	&	\cite{Gilmore84}\\   
		  $UBV$	&	dwarf, giant	&	350	&	3.8	&	\cite{Yamagata92}\\
		  $OE$	&	dwarf, giant	&	325	&	3.2	&	\cite{Larsen96}  \\ 
		  $BVI$	&	dwarf	        &	260	&	2.3	&	\cite{Ojha96}\\
		 $UBV$	&	dwarf, giant & 250-270 &	2.5	&	\cite{Robin96,Robin2000}\\
		 $UGR$	&	dwarf	        &	290	&	4.0	&	\cite{Buser98,Buser99}\\
		 $UBV$	&	dwarf	        &	240	&	2.5	&	\cite{Ojha1999}\\
$u'g'r'i'z'$	&	dwarf, giant	&	330	&	2.2&	\cite{Chen01}\\
CADIS, $B_c$, $R_c$, $I_{815}$&	dwarf	&	267-360	&	1.0-1.5 &\cite{Phleps2000}\\
		$JHK_s$	&	dwarf, giant	&	250	&	2.8	&	\cite{Ojha01}\\
		 $K_s$	&	dwarf, giant	&	285	&	3.3	&	\cite{Lopez02}\\
		$UBVRI$	&	dwarf	        &$280/350$& 2.0-2.5&	\cite{Siegel02}\\
		  $OE$	&	dwarf, giant	&	$-$	&	3.5	&	\cite{LH03}\\
		$JK_s$	&	dwarf, giant	&	268	&	2.1	&	\cite{Cabrera05}\\
		$ugriz$	&	dwarf	        &	220	&	1.9	&	\cite{Bilir06a}\\
		$ugriz$	&	dwarf	        & 220-360 &	1.6-2.5 & \cite{Karaali07}\\
		$ugriz$	&	dwarf	        &	300	&	2.6	&	        \cite{Juric08}\\
		$ugriz$	&	dwarf	        & 325-369	& 1.0-1.9 &	\cite{YK10}\\
	    $K_s$	&	dwarf, giant	&	360	&	3.7	&	\cite{Chang11}\\
		$ugriz$	&	dwarf	        &	256	&	3.5	&	\cite{Bovy12}\\
		$ugriz$	&	giant	        &	$-$	&	3.4	&	\cite{Cheng12}\\
		$JHK_s$	&	dwarf, giant	&	205	&	2.1	& \cite{Polido13}\\
		$ugriz$ &	dwarf	        &	240	&	2.0	&	 \cite{Lopez14}\\
		$JW1$   &	giant	        &	$-$	&	2.0 &	This work\\
		\hline \hline
		\end{tabular}  \label{par}
	\end{center}
\end{table*}

\section{Solar Space Density of the RC Stars}

The RC star sample (the main sample) in the {\it WISE} All-Sky Data Release catalogue of \cite{Cutri13} used for the estimation of the scale-length in the Galactic plane is not complete within distances $0 < r \leq 1.5$ kpc. Here, $r$ is the distance from the Sun. Hence, we used a second sample taken from the re-reduced {\it Hipparcos} catalogue \citep{vanLee07} just to fix the position of the RC stars in the solar neighbourhood. Thus, the solar space density of the second sample would be used to avoid any probable degeneracy in the space density functions of the RC stars in the main sample. 

RC stars consist of a sub-sample of {\em Hipparcos} catalogue which contains about 118 000 objects \citep{vanLee07}. We omitted those with relative parallax errors $\sigma_{\pi}/\pi \leq 0$, so we used 113 596 of them with $\sigma_{\pi}/\pi > 0$ to obtain more accurate data. The estimation of the interstellar extinction is based on the following density law for dust in \cite{Sharma11} which can be reduced to an equation with respect to the Galactic coordinates and distance of a star: 
\begin{eqnarray}
\rho_{dust}(R,z)=\frac{\rho_0}{k_{fl}} \times \exp \left[\frac{R_0-R}{h_R}-\frac{z-z_w}{k_{fl}h_z}\right],
\end{eqnarray}
where $R_0 = 8$ kpc is the distance of the Sun to the Galactic center, $h_R$ ($4.2$ kpc) and $h_z$ ($88$ pc) are the scale-length and the scale-height of the dust, $\rho_0 = 0.54$ mag kpc$^{-1}$, and $k_{fl}$ and $z_w$ describe the flaring and warping of the gas, respectively \citep{Sharma11}:
\begin{eqnarray}
\begin{aligned}
k_{fl}(R) & =1+\gamma_{fl} \times \min(R_{fl} ,R-R_{fl}), \\
z_w(R,\Phi) & =\gamma_{w} \times \min(R_w ,R-R_w) \times \sin\Phi.
\end{aligned}
\end{eqnarray} 
Here, $\gamma_{fl}$, $\gamma_{w}$, $R_{fl}$, and  $R_{w}$ values are $0.0054$, $0.18$, $1.12R_0$ and $8.4$ kpc, respectively. $\Phi$ is the Galactocentric azimuth that increases in the direction of the Galactic rotation and place the Sun at $\Phi = 0$. 

Following the comments of \cite{Arce1999} and \cite{Schlafly2011}, we reduced the colour excess at infinite distance in \cite{Schlegel98}, which will be labeled as $E_{\infty} (B-V)_{initial}$ hereafter, to the final one by the following equation \citep{Binney2014}:
\begin{eqnarray}
\begin{aligned}
E_{\infty} (B-V)_{final}&=E_{\infty} (B-V)_{initial}  \\
&\times C(E(B-V)),
\end{aligned}
\end{eqnarray} 
where,
\begin{eqnarray}
\begin{aligned}
C(&E(B-V)) = 0.6 + 0.2 \\
&\times \left[1-\tanh\left(\frac{E_{\infty} (B-V)_{initial}-0.15}{0.3}\right)\right].
\end{aligned}
\end{eqnarray} 
Then, the colour excess of a star at distance {\em r} can be evaluated by the following equation:
\begin{eqnarray}
E_r(B-V)=E_{\infty} (B-V)_{final} \times \frac{\int_0^r \! \rho(s)ds}{\int_0^{\infty} \! \rho(s)ds},
\end{eqnarray} 
where $\rho (s)$ is the reduced density law in Eq. (1) in terms of distance relative to the Sun, {\em r}, and Galactic coordinates of a given star. 
 
In our calculations, the distance $r$ of a sample star is evaluated by means of its parallax taken from the {\it Hipparcos} catalogue, and the integrals in Eq. (5) are replaces by two sums, i.e.  $\displaystyle \sum_{s=0}^{r} \rho (s)$ and $\displaystyle \sum_{s=0}^\infty \rho (s)$, where the density $\rho (s)$ has been evaluated in steps of 5 pc of the distance. We used the classical equation $ A_V / E(B-V) = 3.1$ for evaluation of the total absorption, $A_V$, and combined it with the $V$ apparent magnitude of the star in question, taken from the \cite{ESA97}, to estimate its absolute magnitude:
\begin{eqnarray}
M_V=V-A_V+5-5\log r.
\end{eqnarray} 

The absolute magnitude-colour diagram of the star sample is plotted in Fig. 1. The position of the RC stars is prominent in this diagram. However, we do not exclude the possibility of contamination of the red giants. As stated in \cite{Nataf14}, there is only a small colour difference between the RC stars and the Red Giant Branch (RGB), i.e. $(V - I)_{RGB}-(V - I)_{RC}\approx$0.15 mag, for star sample with a given metallicity and luminosity. However, we limited the size of the region occupied by the RC stars to decrease the possible contamination of the RGB stars.

We used 9789 of them in the condense region with colour $0.8 \leq (B - V)_0 \leq  1.2$ mag and absolute magnitude $0 \leq  M_V \leq 2$ in our calculations. This constraint is rather useful for our purpose, because our aim in this chapter is to fix a space density for the RC stars in the solar neighbourhood but not to obtain a density distribution for a distance interval. RC stars satisfying this constraint lie within the distances $100 < r \leq 300 $ pc. The space density of these stars has been evaluated by the formula $D(r) = N / V(r)$, where $V(r)$ is the volume between two spheres with radii $r_1 = 100$ and $r_2 = 300$ pc, i.e. $V(r) = \frac{4}{3} \pi (r^3_2-r^3_1)$, and $N$ is the number of RC stars in this volume. The logarithmic value of the space density of the RC stars thus evaluated is $D^* = \log D(r) + 10 = 5.954\pm0.004$.

\section{Scale-length of the Thin Disc in the Galactic Plane}

\subsection{Identification of the RC stars in the $(J,~J-W1)$ CMD}

The RC stars used for the estimation of the scale-length of the thin disc in the Galactic plane are taken from the {\it WISE} catalogue \citep{Cutri13}. We plotted the $(J,J - W1)$ CMD of the stars with Galactic latitudes $-0\overset{^\circ}.5 \leq b \leq +0\overset{^\circ}.5$ in 19 fields with consecutive Galactic longitude intervals, i.e. $85^\circ < l \leq 95^\circ$, $95^\circ < l \leq 105^\circ$, ..., $265^\circ < l \leq 275^\circ$ and identified the RC stars in each field using the Galaxia model \citep{Sharma11} as explained in the following. We applied the Galaxia model to each field defined in our study and obtained the $(J, J - W1)$ CMD for a set of synthetic stars. Then, we adopted the following constrains to separate the RC ones in each field: $1. 8 < \log g~(cgs) \leq 2.6$ and $4500 < T_{eff}~(K) \leq 4900$. As stated in Section 2, the colour difference between the RC stars and RGs is small. Then the same case holds also for the temperatures for two sub-samples. Additionally, \cite{Nataf14} claimed the surface gravity of the Red Giant Branch Bump (RGBB) as log $g\approx 2.6$ (cgs) which is close to that of RC stars we adopted. Hence, we expect a possible contamination of the RG stars in our study.

The CMD for the field with $175^\circ < l \leq 185^\circ$ is plotted in Fig. 2 as an example. It turned out that the right mode in the right panel of this figure corresponds to the RC stars. Then, the same case should hold for the observed RC stars, i.e. the right mode in the left panel consists of RC stars we are interested in, while the left one is occupied by the dwarfs. Although the limiting apparent $J$ magnitude of the RC stars in our study is $J = 17$ mag, we could considered only those with $J\leq$ 14.5 mag which corresponds to $r\leq$ 8 kpc.

Fig. 2 shows that there is a gap between the dwarfs and RC stars with apparent magnitude $J\leq$ 13.5 mag. We could extend the borders separating the two star categories down to the apparent magnitude $J =$ 14.5 mag by checking the consistency of the space densities of the observed RC stars with the ones of the synthetic RC stars, for a given field.

We applied the constraint ``data quality flags" to the {\em WISE} catalogue \citep{Cutri13} to select the best possible data. Hence, we expect the best possible data. Hence, we expect the possible best result in scale-length estimation of the thin disc in our study.

\subsection{Estimation of the Scale-length of the Thin Disc}

The scale-lengths of the thin disc in the Galactic plane have been estimated by fitting the observed space density functions of RC stars in 19 fields to the Galaxy models. The distances of the RC stars relative to the Sun used in the density functions have been evaluated as follows. The $E(J-W1)$ colour excess of a RC star has been adopted as the difference between its observed $J - W1$ colour index and the mean true colour index $(J-W1)_0 = 0.665\pm 0.032$ mag of the RC sample \citep{Yaz13}, while the equation for the total absorption, $A_J = 1.217 \times E(J-W1)$, is adopted from \cite{Bilir2011}. Then, the distance {\em r} could be calculated from the following well known 
formula:
\begin{eqnarray}
J-M_J-A_J=5\log r-5,
\end{eqnarray} 
here, $M_J = - 0.970 \pm 0.016$ mag \citep{Yaz13}. 

We assumed that the RC stars in the Galactic latitude range $-0\overset{^\circ}.5 \leq b \leq +0\overset{^\circ}.5$ are all thin disc stars. Hence, the logarithmic space densities $D^* = \log D + 10$ evaluated for 19 star fields belong only to the thin disc population of our Galaxy. Here, $D = N / \Delta V_{1,2}$; $\Delta V_{1,2} = (\frac{\pi}{180})^2 \frac{A}{3}(r^3_2-r^3_1)$; $A$ denotes the size of the field $(10$ deg$^2)$; $r_1$ and $r_2$ are the lower and upper limiting distances of the volume $\Delta V_{1,2}$; $N$ is the number of RC stars in this volume; $r^* = [(r^3_1+r^3_2)/2]^{1/3}$ is the centroid distance of the volume $\Delta V_{1,2}$. 

We adopted the density law of Basle group \citep{Buser98,Buser99} for the thin disc. Disc structures are usually parametrized in cylindrical coordinates using radial and vertical exponentials:
\begin{eqnarray}
D(x,z)=n \times \exp\left(\frac{-|z|}{H}\right) \times \exp\left(\frac{-(x-R_0)}{h}\right),
\end{eqnarray}
where $ z = z_\odot + r \sin b$, $r$ is the distance from the Sun, $b$ is the Galactic latitude, $z_\odot$ is the vertical distance of the Sun from the Galactic plane (25 pc; \citealp{Juric08}), $x$ is the projection of the Galactocentric distance on the Galactic plane, $R_0$ is the solar distance from the Galactic centre (8 kpc), $H$ and $h$ are the scale-height and scale-length, respectively, and $n$ is the normalized local density at the solar radius. In our case, we assume that the space density does not vary in the vertical direction due to the small range of the fields, i.e. $-0\overset{^\circ}.5 \leq b \leq +0\overset{^\circ}.5$. Then, assuming that $\exp(-|z|/H)\approx1$, Eq. (8) can be reduced to the following form:
\begin{eqnarray}
D(x)=n \times \exp\left(\frac{-(x-R_0)}{h}\right).
\end{eqnarray} 

We fitted the space density functions derived from the observations to the density law adopted for the thin disc and estimated the model parameters {\em n} and {\em h} for each field. We used the classical $\chi_{min}^2$ statistic for the estimation of the model parameters \citep{Phleps2000,Phleps2005,Chen01,Siegel02,Du2003,Du2006,Juric08,Ak07,Bilir08,YK10}. The comparison of the logarithmic density functions derived from the observations and the analytical density law for the thin disc is given in Fig. 3 or four fields with Galactic longitude $l = 90^\circ, 140^\circ,180^\circ$ and $260^\circ$, as example, while the results are tabulated in Table 2 for all fields. The errors have been evaluated by changing a Galactic model parameter until an increase or decrease in $\Delta \chi^2$  by 1 was achieved \citep{Phleps2000}. 

We used the synthetic RC stars identified by the Galaxia model \citep{Sharma11} and estimated a second set of Galactic model parameters which are also given in Table 2. Also, the logarithmic density functions for the same fields, $l = 90^\circ, 140^\circ,180^\circ$ and $260^\circ$, are presented in Fig. 3 as example, as well as for comparison with the corresponding ones of observed RC stars. Table 2 shows that the solar normalizations estimated for two sets of data, i.e. for the observed RC stars and for the synthetic ones, for a given field are compatible. Also, they confirm the solar space density, $D^*(0) = 5.954$, estimated in Section 2 by means of {\it Hipparcos} RC stars. We plotted the scale-lengths estimated for the same data in Fig. 4, versus Galactic longitudes. Table 2 and Fig. 4 show also the agreement of the scale-lengths corresponding two sets of data for a given longitude.

 \begin{table*}[t]
 	\centering
 	\caption{Galactic model parameters for the thin disc estimated by using  the observed RC stars in {\em WISE} and the synthetic ones in Galaxia for 19 star fields. $N$ indicates the number of star for a given field.}
 		\begin{tabular}{ccccc|cccc}
 			\hline\hline
 			& \multicolumn{4}{c}{Observed data} & \multicolumn{4}{c}{Synthetic data} \\
 			\hline
$l(^\circ)$  &      $N$      &  $D^*$                &  $h$ (pc)             &   $\chi^2_{min}(\times10^{-6})$  &   $N$   &  $D^*$ &   $h$ (pc)              &   $\chi^2_{min}(\times10^{-6})$  \\
\hline
$90$   &   $15693$   &   $6.01^{+0.02}_{-0.03}$   &   $1597^{+109}_{-137}$   &   $3.3$   &   $15852$   &   $5.99^{+0.03}_{-0.02}$   &   $1635^{+140}_{-88}$   &   $2.3$\\
$100$   &   $11318$   &   $5.99^{+0.04}_{-0.05}$   &   $1624^{+117}_{-127}$   &  $11.5$   &   $12310$   &   $5.98^{+0.10}_{-0.10}$   &   $1634^{+306}_{-146}$   &   $22.6$\\
$110$   &   $7911$   &   $6.04^{+0.03}_{-0.08}$   &   $1574^{+94}_{-141}$   &   $10.8$   &   $8817$   &   $6.04^{+0.07}_{-0.07}$   &   $1579^{+151}_{-84}$   &   $6.7$\\
$120$   &   $6228$   &   $6.00^{+0.04}_{-0.03}$   &   $1697^{+74}_{-69}$   &   $3.5$   &   $6889$   &   $6.08^{+0.08}_{-0.08}$   &   $1575^{+159}_{-93}$   &   $11.5$\\
$130$   &   $5518$   &   $6.00^{+0.06}_{-0.03}$   &   $1797^{+121}_{-66}$   &   $3.9$   &   $5718$   &   $6.08^{+0.07}_{-0.07}$   &   $1667^{+132}_{-79}$   &   $9.7$\\
$140$   &   $5114$   &   $6.04^{+0.05}_{-0.03}$   &   $1911^{+123}_{-68}$   &   $2.2$   &   $4836$   &   $6.06^{+0.04}_{-0.03}$   &   $1814^{+85}_{-58}$   &   $2.6$\\
$150$   &   $4923$   &   $6.00^{+0.04}_{-0.03}$   &   $2021^{+104}_{-68}$   &   $2.3$   &   $4593$   &   $6.03^{+0.06}_{-0.06}$   &   $1896^{+124}_{-76}$   &   $4.2$\\
$160$   &   $4350$   &   $6.02^{+0.07}_{-0.03}$   &   $1956^{+138}_{-61}$   &   $4.6$   &   $4429$   &   $5.99^{+0.07}_{-0.07}$   &   $2036^{+148}_{-78}$   &   $5.1$\\
$170$   &   $4338$   &   $6.03^{+0.06}_{-0.03}$   &   $1998^{+128}_{-67}$   &   $5.5$   &   $4325$   &   $6.04^{+0.03}_{-0.03}$   &   $2003^{+63}_{-51}$   &   $1.9$\\
$180$   &   $3945$   &   $5.98^{+0.04}_{-0.02}$   &   $2137^{+114}_{-58}$   &   $2.3$   &   $4014$   &   $6.03^{+0.05}_{-0.03}$   &   $2044^{+110}_{-67}$   &   $4.0$\\
$190$   &   $4528$   &   $6.02^{+0.06}_{-0.05}$   &   $2012^{+118}_{-76}$   &   $8.5$   &   $4616$   &   $6.01^{+0.05}_{-0.05}$   &   $2026^{+80}_{-65}$   &   $3.9$\\
$200$   &   $4061$   &   $6.01^{+0.06}_{-0.03}$   &   $1926^{+113}_{-58}$   &   $3.3$   &   $4178$   &   $6.05^{+0.05}_{-0.05}$   &   $1887^{+98}_{-57}$   &   $3.3$\\
$210$   &   $4369$   &   $6.05^{+0.06}_{-0.04}$   &   $1895^{+133}_{-68}$   &   $4.9$   &   $4085$   &   $6.05^{+0.05}_{-0.05}$   &   $1846^{+102}_{-51}$   &   $3.1$\\
$220$   &   $5242$   &   $6.07^{+0.04}_{-0.03}$   &   $1817^{+84}_{-54}$   &   $2.1$   &   $4646$   &   $6.07^{+0.07}_{-0.07}$   &   $1697^{+126}_{-77}$   &   $5.6$\\
$230$   &   $5149$   &   $6.05^{+0.05}_{-0.09}$   &   $1712^{+113}_{-151}$   &   $9.7$   &   $5500$   &   $6.11^{+0.05}_{-0.05}$   &   $1604^{+96}_{-68}$   &   $4.4$\\
$240$   &   $6082$   &   $6.04^{+0.06}_{-0.10}$   &   $1598^{+131}_{-131}$   &   $12.3$   &   $6659$   &   $6.10^{+0.08}_{-0.08}$   &   $1489^{+115}_{-89}$   &   $7.5$\\
$250$   &   $6784$   &   $5.99^{+0.05}_{-0.09}$   &   $1549^{+119}_{-153}$   &   $9.5$   &   $8342$   &   $6.05^{+0.07}_{-0.07}$   &   $1524^{+140}_{-83}$   &   $7.7$\\
$260$   &   $10368$   &   $5.97^{+0.03}_{-0.06}$   &   $1577^{+111}_{-143}$   &   $9.0$   &   $11290$   &   $5.98^{+0.07}_{-0.07}$   &   $1546^{+186}_{-102}$   &   $9.9$\\
$270$   &   $15986$   &   $5.93^{+0.02}_{-0.02}$   &   $1633^{+112}_{-78}$   &   $2.8$   &   $16995$   &   $5.97^{+0.04}_{-0.03}$   &   $1566^{+157}_{-125}$   &   $5.2$\\
 			\hline\hline
 		\end{tabular}

 \end{table*}

The distribution of the scale-length for the thin disc in Fig. 4 is not flat. It makes a peak, $h = 2$ kpc, at the Galactic longitude $l=180^\circ$ and it decreases symmetrically in the Galactic longitude intervals $90^\circ \leq l < 180^\circ$ and $180^\circ < l \leq 270^\circ$, reaching to a minimum value of $h=1.6$ kpc at the terminal Galactic longitudes, $l = 90^\circ$ and $l=270^\circ$.   

\section{Discussion}

We estimated the scale-length of the thin disc by means of the space density functions of the RC stars in 19 star fields with consecutive Galactic longitudes, $l = 90^\circ, 100^\circ, ..., 270^\circ$ and with Galactic latitudes $-0\overset{^\circ}.5 \leq b \leq +0\overset{^\circ}.5$. We used the solar space density of the RC stars in the {\it Hipparcos} catalogue, $\log D(0) + 10 = 5.954\pm0.004$, evaluated in this study in fitting the observed space density functions to the Galaxy models. We used the data in the {\it WISE} catalogue \citep{Cutri13} and identified the RC stars by comparison their positions in the $(J, J - W1)$ CMDs with the synthetic ones in Galaxia (private communication and \citealp{Sharma11}). The RC stars are adopted as the objects which satisfy the following conditions: i) $1.8 < \log g~(cgs) \leq 2.6$ and ii) $4500 < T_{eff}~(K) \leq 4900$. Then, the right mode in the $(J, J - W1)$ CMD consists of RC stars, while the left one belongs to the dwarfs (Fig. 2). We used the gap between the dwarfs and RC stars with apparent magnitudes $J \leq 13.5$ mag as a clue and separated these star categories down to the apparent magnitude $J = 14.5$ mag by checking the consistency of the space densities of the observed RC stars with the ones of the synthetic RC stars, for a given field. 

The scale-lengths estimated for the thin disc with the RC stars in different fields are not equal to each other. The largest scale-length, $h = 2137_{-58}^{+114}$ pc, corresponds to the Galactic longitude $l = 180^\circ$. The scale-lengths estimated for the fields with Galactic longitudes close to $l = 180^\circ$, i.e. $l = 150^\circ, 160^\circ, 170^\circ, 190^\circ$ are rather close to $h = 2137_{-58}^{+114}$ pc. However, they deviate from this value systematically in the fields with Galactic longitudes $l \le 150^\circ$ and $l > 190^\circ,$ and they make two flat distributions at the two terminal longitudes, $l = 90^\circ$ and  $l = 270^\circ$. The smallest scale-length, $h = 1574_{-141}^{+94}$ pc, in the second quadrant belongs to the field with Galactic longitude $l = 110^\circ$, while in the third quadrant it corresponds to $l = 250^\circ$, $h = 1549_{-153}^{+119}$ pc. The distribution of the observed scale-length is confirmed by the synthetic RC stars, hence it is real. We estimated the scale-length of the thin disc for three fields, $l = 90^\circ, 180^\circ$, and $270^\circ$ by adopting a scale-height, $H=350$ pc \citep{Siegel02} and compared them with the ones in Table 2. The differences for two sets of data are less than 0.1 kpc for three fields, which are within the errors given in Table 2.

The difference between the scale-lengths estimated for different longitudes can be explained as follows. The first reason is that in the Galactic anticentre direction, we reach in average higher values of Galactocentric distance ($R$). A second  explanation can be done by the overdensities in the direction perpendicular to the Galactic anticentre direction or close to this direction, such as the fields with Galactic longitudes $l = 90^\circ, 100^\circ, 110^\circ, 120^\circ, 240^\circ, 250^\circ, 260^\circ, 270^\circ$. The scale-lengths in the fields with such Galactic longitudes are relatively small, while small number of stars in the fields in the anticentre direction exhibit larger scale-lengths ({Table 2). Additionally, Fig. 5 shows that the distance histograms for three Galactic longitudes, $l = 90^\circ, 180^\circ$, and $270^\circ$ are different. The one for the longitude $l = 180^\circ$ is flat, while the number of RC stars for the Galactic longitudes $l = 90^\circ$ and $270^\circ$ are monotonously increasing with distance. As the volume increases with distance, the flat distribution results a decreasing space density for the Galactic longitude $l = 180^\circ$ (Fig. 3c), while the increasing number of stars for the Galactic longitudes $l = 90^\circ$ and $270^\circ$ supply relatively flat space density functions (Fig. 3a). 

The scale-length for the thin disc appeared in the literature lies in the range $2 \leq h \leq 4$ kpc, the least ones being those of \cite{Robin96,Ojha1999,Robin2000,Chen01,Siegel02}, i.e. $2 \leq h \leq 2.5$ kpc. The most recent scale-length is claimed by \cite{Lopez14}, $h = 2$ kpc. Their observations were carried out in an off-plane low Galactic regions with size $1400$ deg$^2$ and they were fitted to a model of flared thin and thick disc. If we normalize the scale-length, estimated for the RC stars in the fields in the opposite direction of the Galactic centre or close to this direction, to one decimal we obtain exactly the same value, $h = 2$ kpc. This is a good result which can be used as an argument to claim that the value $h = 2$ kpc can be adopted as a standard scale-length for the thin disc for the regions of the Galaxy which are not contaminated by the accreted material or other features such as warp and flare.

In this study, we did not consider the effect of the contamination of the RGs. However, this contamination shows its effect more or less in the studies related to the RC stars due to several reasons. First, the RC stars and the RGB fall in close colour intervals, i.e. the difference between their intrinsic colours in $(V-I)_0$ is only $\approx$ 0.15 mag \citep{Nataf14} which complicates their separation due to the errors in colour estimation and interstellar reddening. Second, the number of RC stars is larger than the one of RGs in a given sample, and their ratio is an increasing function of metallicity. \cite{Nataf14} give this ratio as 1.75 and 2.32 for the metallicities $[M/H]= -1.27$ and $[M/H]= 0.40$ dex, respectively. Finally, as cited in \cite{Nataf14} the fainter RC stars and the RGBB overlap at the distance modulus 14.80 mag (their Figure 2).

In our study the probable effect of the first reason is unvoiddable, while we expect minimum effect due to the second reason, because our RC sample covers only the disc stars with solar metallicity in the Galactic plane. The distance modulus 14.80 mag just mentioned corresponds to the apparent magnitude $J_0 \approx 13.80$ mag in our study which indicates that only the faintest stars, $13.50 < J_0 < 14.50$ mag, can be affected from the overlapping in question. However, the faint segments of our space density functions show a smooth extension relative to the bright ones. Hence, we do not expect a large contamination due to this reason. All these statements are valid for the Galaxia data, because they are based on a set of atmospheric parameters which can be valid for RC stars as well as RGs. 

{\bf Conclusions:} The scale-length of the thin disc estimated with the most probable RC stars in the Galactic plane is 2 kpc for the fields in the Galactic anticentre direction or close to this direction, while it decreases continuously in the second and third quadrants reaching to a lower limit of $h = 1.6$ kpc at the Galactic longitudes $l= 90^\circ$ and $l= 270^\circ$.

\section{Acknowledgments}
 We would like to thank anonymous referee for the useful comments and corrections. Also, we would like to thank Dr. Sanjib Sharma for providing an updated version of Galaxia model. This study has been supported in part by the Scientific and Technological Research Council (T\"UB\.ITAK) 112T120. Part of this work was supported by the Research Fund of the University of \.Istanbul, Project Number Numbers: 48483. MLC was supported by the Spanish Ministry of Economy and Competitiveness (MINECO) under the grant AYA2012-33211. This research has made use of NASA's Astrophysics Data System Bibliographic Services, the VizieR catalogue access tool, CDS, Strasbourg, France and NASA/IPAC Infrared Science Archive and Extragalactic Database (NED), which are operated by the Jet Propulsion Laboratory, California Institute of Technology, under contract with the National Aeronautics and Space Administration. This publication makes use of data products from the Wide-field Infrared Survey Explorer and the Two Micron All Sky Survey, which are a joint project of the University of Massachusetts and the Infrared Processing and Analysis Center/California Institute of Technology, funded by the National Aeronautics and Space Administration and the National Science Foundation.

\begin{figure*}
\centering
\includegraphics[width=\textwidth]{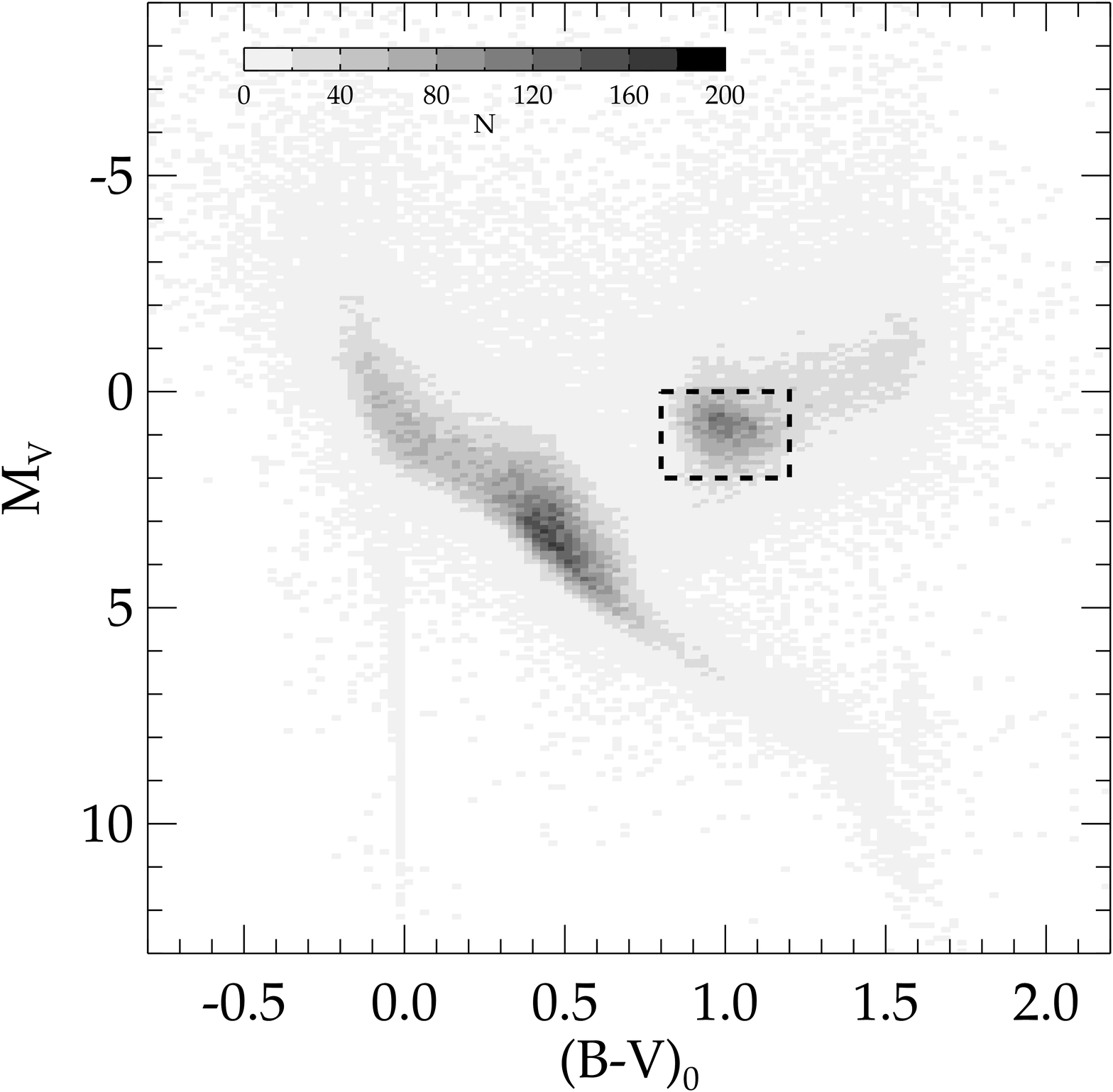}
\caption{Absolute magnitude-colour diagram for the stars in the re-reduced {\em Hipparcos} catalogue. The location of the most probable RC stars is indicated with a rectangular.} 
\end {figure*} 
\pagebreak

\begin{figure*}
\centering
\includegraphics[width=\textwidth]{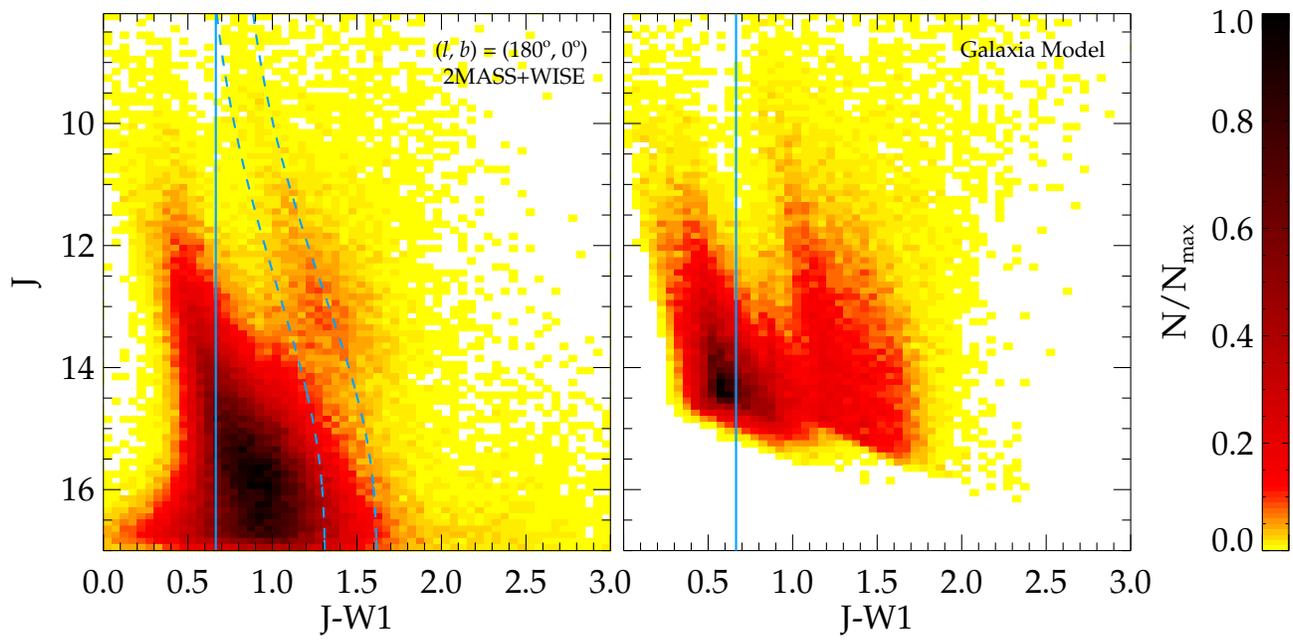}
\caption{Colour-magnitude diagrams for the observed (left panel) and synthetic (right panel) stars in the field with Galactic longitude $175^\circ < l \leq 185^\circ$ and Galactic latitude $-0\overset{^\circ}.5 \leq b \leq +0\overset{^\circ}.5$. Blue solid lines show intrinsic colour, $(J-W1)_0=0.665$ mag of RC stars \citep{Yaz13}, light blue-dashed lines display the borders of the most probable RC stars.} 
\end {figure*} 
\pagebreak
		
\begin{figure*}
\centering
\includegraphics[width=\textwidth]{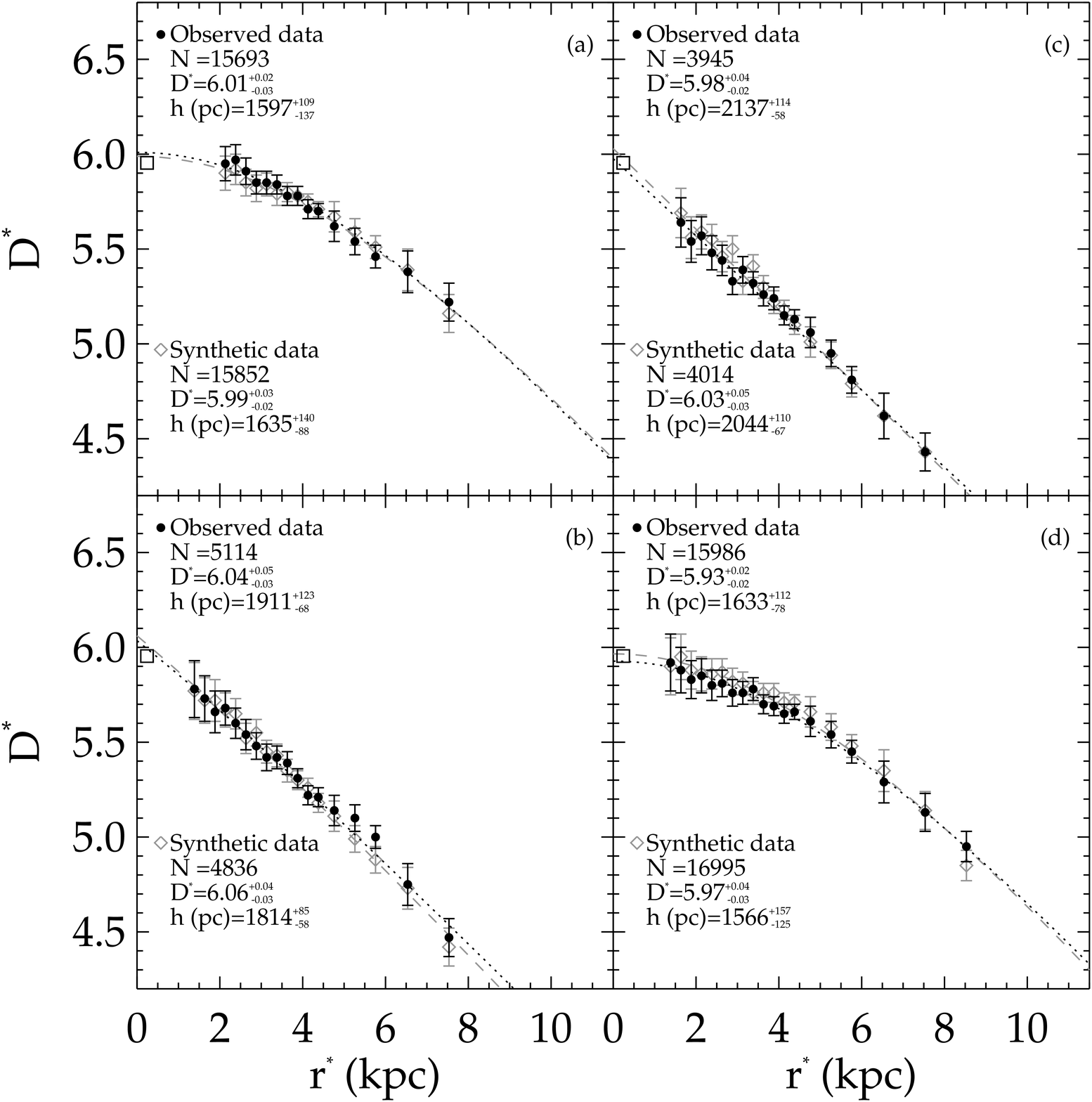}
\caption{Logarithmic space density functions for the most probable RC stars in the fields with Galactic longitudes (a) $l = 90^\circ$, (b) $140^\circ$, (c)  $180^\circ$, (d) $270^\circ$. Filled circles indicate the observed data, while the open ones correspond to the synthetic data in Galaxia.} 
\end {figure*} 
\pagebreak
		
\begin{figure*}
\centering
\includegraphics[width=\textwidth]{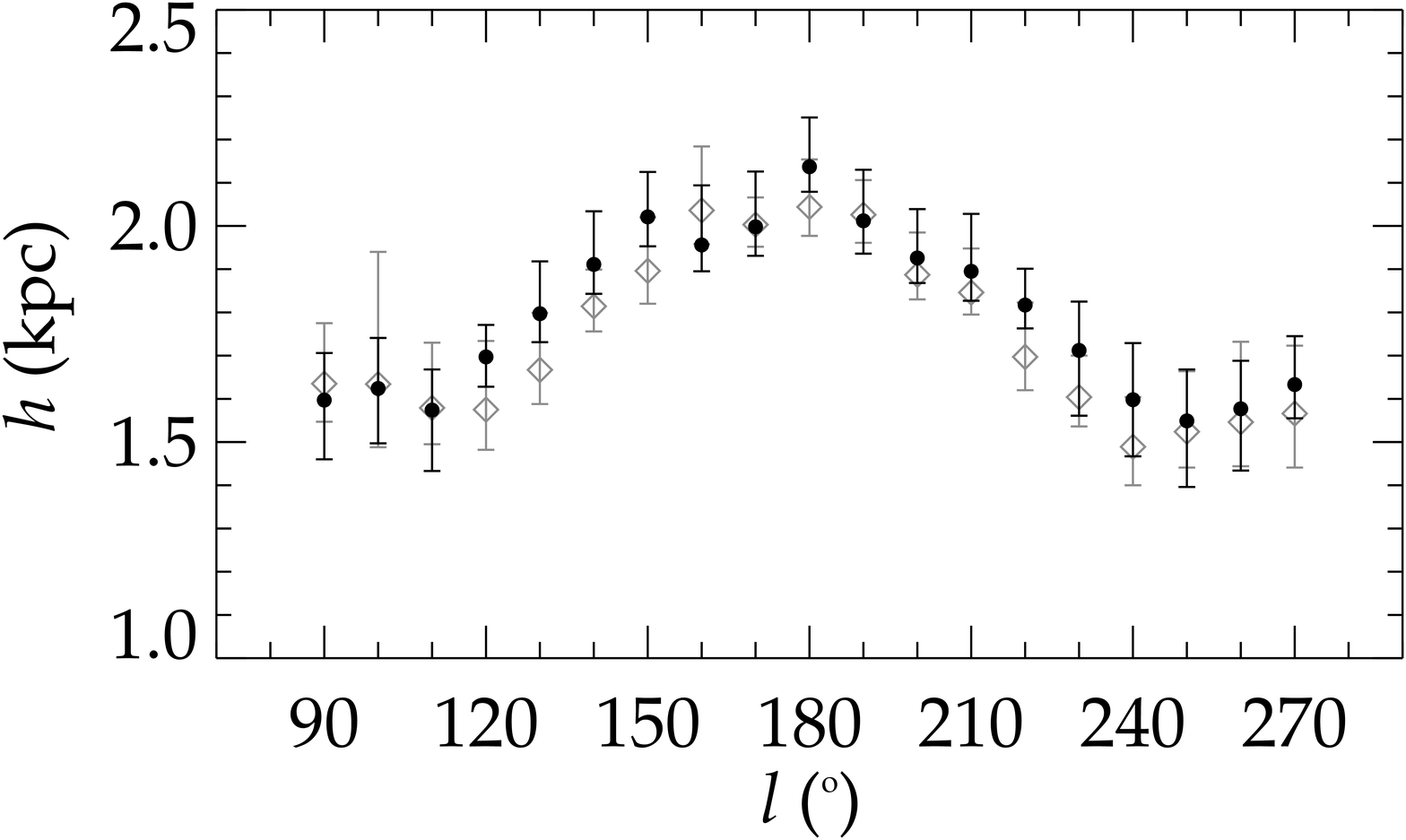}
\caption{Scale-lengths for the observed ($\bullet$) and synthetic ($\diamond$) data versus Galactic longitude.} 
\end {figure*} 
\pagebreak
	
\begin{figure*}
\centering
\includegraphics[width=0.75\textwidth]{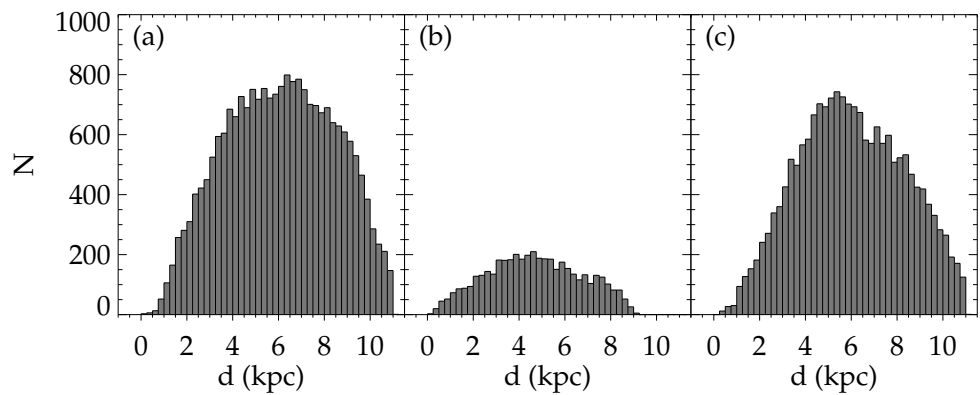}
\caption{Distance histograms for the most probable RC stars in three different star fields: (a) $l=90^\circ$, (b) $l=180^\circ$, and (c) $l=270^\circ$.} 
\end {figure*} 
\pagebreak

\end{document}